\documentclass[twocolumn]{aastex63}

\usepackage{amsmath}
\usepackage{wasysym}
\usepackage{hyperref}
\usepackage{xcolor}
\usepackage{url}
\usepackage{longtable}
\RequirePackage{lineno}

\newcommand{\YQ}{$Y_{\textrm{Q}}$}
\newcommand{\XQ}{$X_{\textrm{Q}}$}
\newcommand{\AQ}{$A(\textrm{Q})$}
\newcommand{\sol}{$_{\odot}$}

\def\gtaprx{ \mathrel{ \vcenter{
      \offinterlineskip \hbox{$>$}
      \kern 0.3ex \hbox{$\sim$}    } } }

\def\ltaprx{ \mathrel{ \vcenter{
      \offinterlineskip \hbox{$<$}
      \kern 0.3ex \hbox{$\sim$}    } } }

\def\apjs{{ApJS}}
\def\apjl{{ApJL}}

\def\aap{{A\&A}}

\shorttitle{Converting Stellar Z to Abundances to Molar Ratios}
\shortauthors{Hinkel, Young, \& Wheeler}

\begin{document}

\title{A Concise Treatise on Converting Stellar Mass Fractions to Abundances to Molar Ratios}

\author[0000-0003-0595-5132]{Natalie R.\ Hinkel}
\affiliation{Southwest Research Institute, 6220 Culebra Rd, San Antonio, TX 78238, USA}

\author[0000-0003-1705-5991]{Patrick A. Young}
\affiliation{School of Earth and Space Exploration, Arizona State University, Tempe, AZ 85287, USA}

\author[0000-0001-5563-6987]{Caleb H. Wheeler III}
\affiliation{Simons Foundation, Center for Computational Astrophysics, 162 5th Ave, New York, NY, 10010, USA}

\correspondingauthor{Natalie Hinkel}
\email{natalie.hinkel@gmail.com}

\begin{abstract}

Understanding stellar composition is fundamental not only to our comprehension of the galaxy, especially chemical evolution, but it can also shed light on the interior structure and mineralogy of exoplanets, which are formed from the same material as their host stars. Unfortunately, the underlying mathematics describing stellar mass fractions and stellar elemental abundances is difficult to parse, fragmented across the literature, and contains vexing omissions that makes any calculation far from trivial, especially for non-experts. In this treatise, we present clear mathematical formalism and clarification of inherent assumptions and normalizations within stellar composition measurements, which facilitates the conversion from stellar mass fractions to elemental abundances to molar ratios, including error propagation. We also provide an example case study of HIP 544 to further illustrate the provided equations. Given the important chemical association between stars, as well as the interdisciplinary relationship between stars and their planets, it is vital that stellar mass fractions and abundance data be more transparent and accessible to people within different sub-fields and scientific disciplines.

\end{abstract}

\keywords{stellar abundances; high resolution spectroscopy; planetary structure; meteorite composition; solar abundances; interdisciplinary astronomy}

\section{Motivation}
In the early universe, the only elements created in any notable amount were hydrogen, helium, and to a lesser extent lithium. With the formation of the first stars, where interior temperatures and pressures reached levels for nuclear fusion, the lighter elements combined to form heavier atoms \citep[][and references therein]{Arnett96}. This first generation of extremely massive stars, $\sim$100 M$_{\odot}$, ended their lives by exploding, sending out the newly created heavy elements (referred to as ``metals" by astronomers) that diversified the chemical population of the universe. The next generation of stars were formed not only from the Big Bang H and He, but they also contained some of the heavy elements, which they continued to fuse during their lifetimes. It was through the cycles of multiple generations of stars that most of the elements in the Periodic Table were formed. Since, stars and planets are created at the same time, from the same molecular birth cloud, the chemistry of planets is intrinsically linked to that of their host stars \citep[e.g.][]{Gonzalez97, Fischer05, Ramirez09, Melendez09, Mulders18}. 

The field of exoplanet science is inherently interdisciplinary. In most cases, nearly everything that we know about a given exoplanet is inferred from the way that it interacts with its host star. For example, understanding the way in which the planet gravitationally influences the star (e.g. via the radial velocity technique) provides a means for the calculation of planetary mass \citep[e.g.][and references therein]{Wright18}. The planet radius may be determined by comparing the size of the planet to the host star during a transit observation \citep[e.g.][]{Kane09, Deeg18}. Similarly, the composition of the exoplanet atmosphere can be determined from how the light of the host star is transmitted or absorbed as it passes through a planet's exosphere, such as with JWST\footnote{\url{www.jwst.nasa.gov/}} \citep[][and references therein]{Madhusudhan18, Kreidberg18}. 

Smaller rocky planets are not directly resolvable from their host stars and they contribute a comparatively insignificant amount of light. Therefore, it is not currently possible to measure the interior structure or solid surface composition of rocky planets. However, the amount of various elements within the star, or elemental abundances, can be used as a logical 1:1 proxy when modeling the interior make-up and mineralogy of a small planet \citep[e.g.][]{bond_2010_aa, Thiabaud15, Hinkel18, Schulze21}. 

Stars are made up of shells of nuclear fusion; these ``burning" shells transport energy via convection or radiation. While interior shells are observationally opaque, it is from the outer layers of the star -- or the photosphere -- that light is observable \citep[][and references therein]{Arnett96}. It is therefore predominantly the photosphere that is measured when determining the stellar composition. Since the photosphere does not experience the temperatures and pressures required for burning, it is assumed to be similar in composition to the proto-solar cloud from which the star formed. However, there are a few processes which may impact the abundances over time: heavy elements in the photosphere will settle into the interior over the course of the star's lifetime \citep[or gigayears, e.g.][]{Dotter17}, radioactive isotopes will decay, and nuclear processing may destroy fragile elements (Li, Be, and B) mixed from the stellar surface by convection. Elemental abundances are therefore indirectly measured by using realistic models of the physics within the stellar photosphere. High accuracy atomic and molecular line data, usually determined within a laboratory, are also necessary for identifying lines within the stellar spectrum. 

To take advantage of all of the ways in which stars and planets may interact, scientists from a variety of backgrounds need to be able to convert measurements and units across disciplines. Here we peel back the stellar composition notation to clearly outline how stellar hydrogen ($X$), helium ($Y$), and the remaining heavier elements ($Z$), are defined by stellar theorists. We discuss how to convert from stellar $XYZ$ mass fractions to elemental abundance ratios to more geologically useful units, namely molar ratios. While there have been partial, piecemeal explanations of stellar composition in the literature \citep[e.g.][]{Hinkel14, Roederer18, Hinkel20, Gebruers21}, there does not exist a clear breakdown of the notation and conversion making the compositional relationships between stars and also their planets more difficult to decipher. More importantly, this limits the tractability, reproducibility, and accessibility of stellar composition data for astrophysicists outside of the stellar abundance sub-field and for scientists as a whole.

In this paper, we provide a rigorous mathematical explanation of the many ways in which stellar composition may be presented by either theorists or observers. In \S \ref{s.math}, we go over the deep-rooted formalism within stellar elemental abundances, outlining a clear mathematical notation while also providing alternatives found within the literature. In \S \ref{s.massfractions}, we explain the definition of stellar mass fractions and the underlying assumptions associated with those terms.  In \S \ref{s.abundstomoles}, we show generally how to determine molar ratios from stellar abundances and demonstrate how to propagate errors. In \S \ref{s.examplecase}, we illustrate the equations outlined in \S \ref{s.massfractions} and \S \ref{s.abundstomoles} by providing a step-by-step walk through of an example case study of HIP 544, including a table of calculated values for each step. In light of the solar normalization associated with the stellar abundance notation, in \S \ref{s.solar_abunds} we go over key aspects necessary for solar abundance determination and briefly describe the variations between popular measurements. Finally, in \S \ref{s.discussion} we provide a discussion and go over major caveats when working with stellar composition, such as observing elements in different ionization states and the application of non-local thermodynamic equilibrium (NLTE) to stellar abundance prescriptions.

\section{Mathematical Formalism}\label{s.math}
Most observed stellar elemental abundance values are reported in dex, or ``decadic logarithmic unit" \citep{Lodders19}. Logarithmic scales (metrics) are useful for tracking observables that span large dynamic ranges and/or when observables are more naturally represented as ratios in linear units, which then become simple addition and subtraction of operations in log-space. The ``decadic'' of dex tells us that our log base is 10, and not the natural log often seen in other applications. Another common base-10 log unit commonly seen is the decibel (dB). While dB scale is defined absolutely with 1 W = 0 dB, the relative span of the units can be simply compared to dex as the change in $\Delta 1$ dex$ = \Delta10$ dB.

While we define dex as a unit on a base-10 log scale, we distinguish dex-abundance-notation as an additional formalism that includes absolute scaling (i.e., to hydrogen, as discussed in \S \ref{s.hnorm}) and defines allowed operations on abundance values. However, abundance literature rarely makes this disambiguation, leaving the reader to determine from context whether the unit or the abundance formalism is being discussed.

Dex-abundance-notation includes numerous expressions to indicate comparison, for example: brackets to indicate inherent solar normalization (see \S \ref{s.solarnormmath}), e.g. [Q/H] or [Q/Fe]. While this notation has historical precedent and utility to experts, the reported measurements are often difficult for those outside of the sub-field to interpret, distinguish, or convert to other quantities. Here we explain the underlying meaning and mathematical framework for stellar abundances.

To begin the formalism of dex-abundance-notation, we summarize variables and definitions. Abundance number fractions (not to be confused with more geologically-driven molar ratios discussed in \S \ref{s.abundstomoles}) have a scale which ranges from [0, 1], where 0 means this element is not present and 1 meaning that an object is composed entirely of that element, such that the sum of all number fractions within an object add to 1. The letter `q' is used to represent a number fraction for a generalized element and the letter `h' is used to represent the number fraction of hydrogen. Similarly, `fe' represents the number fraction of iron and is used as a specific example of a generalized element, other than h and q, to demonstrate scaling by that element.

\subsection{Scaling to Hydrogen}\label{s.hnorm}
The logarithmic and relative scale of dex-abundance-notation offers a quick way to compare across elements and between stars. In the following section we demonstrate the process for going from number fractions for an element to the relative scale of dex-abundance-notation, which makes comparisons simple by using subtractions in log-space. 

Astronomers define the abundance of an element starting from a ratio with respect to the number of hydrogen atoms fixed at trillion \citep[or 10$^{12}$ per][]{Payne25a, Payne25b, Claas51}. For a generic element $\textrm{Q}$, let $\mathcal{Q}$ represent the number of atoms in a star per every (10$^{12}$) hydrogen atoms. Written in terms of number fractions, we have 

\begin{equation}\label{e_molfract}
    \mathcal{Q} = \frac{\textrm{q}}{\textrm{h}} \times 10^{12}.
\end{equation}

\noindent
For the element $\textrm{Q}$ we define the absolute stellar abundance, $A(\textrm{Q})$, as

\begin{equation}\label{e_q_abs_abundance}
    A(\textrm{Q}) \equiv  \log_{10}(\mathcal{Q})\,\,,
\end{equation}

\noindent
such that the two quantities are mathematically equivalent ($\equiv$) to each other. Combining Eqs. \ref{e_molfract} and \ref{e_q_abs_abundance} for the case of $\textrm{q}=\textrm{h}$ shows that the absolute stellar abundance of hydrogen, represented by $\textrm{H}$, as 
\begin{equation}\label{e_h_abs}
\begin{split} 
   A(\textrm{H}) &= \log_{10}(\mathcal{}H) \\
   &= \log_{10}\Big(\frac{\textrm{h}}{\textrm{h}} \times 10^{12} \Big) \\
    &= \log_{10}(10^{12}) \\
    &=12.\\
\end{split}
\end{equation}

\noindent
Hydrogen is the dominant element for all stars, thus the upper bound for the absolute abundance of elements, $A(\textrm{Q})$, in the dex-abundance-notation is 12 and is equal to the absolute abundance of hydrogen. There is no lower limit to the dex-abundance-notation and negative values are allowed. Although absolute abundances in dex-abundance-notation are, for the most part, positive. Written in a purely mathematical sense, the dex scale for absolute abundances can be any number in the range $(-\infty, 12]$, however there are practical limits on the measurement of sufficiently rarefied elements.

Because hydrogen is the most prevalent element for all stars, we can formally state that the number fraction for h is nearly equal to 1, which we write as:

\begin{equation}\label{e_h_molfract}
    \textrm{h} \approx 1\,\,.
\end{equation}

\noindent
Eq. \ref{e_h_molfract} leads to the simplification of Eq. \ref{e_molfract}, namely

\begin{equation}\label{e_molfract_simple}
    \mathcal{Q} \approx \textrm{q} \times 10^{12} \\
\end{equation}

\noindent
With Eq. \ref{e_molfract_simple}, we are now prepared to write Eq. \ref{e_q_abs_abundance} in terms of the number fraction, q, and the definition of the absolute abundance of hydrogen in Eq. \ref{e_h_abs} as 

\begin{equation}\label{e_q_abs}
\begin{split} 
    A(\textrm{Q}) &=  \log_{10}(\textrm{q} \times 10^{12})\\
       &= \log_{10}(\textrm{q}) + 12 \\
       &= \log_{10}(\textrm{q}) + A(\textrm{H}). \\
\end{split}
\end{equation}

\noindent
Note that Eq. \ref{e_q_abs} contains a factor of $10^{12}$ required to scale a number fraction of unity to an absolute stellar abundance $12$, as is the case for hydrogen. In addition, because the logarithm of any number fraction value $<$ 1 is negative, that means that any non-hydrogen A(Q) will be less than 12. To acknowledge that \AQ\, is defined in terms of $\textrm{H}$, dex-abundance-notation defines absolute abundance with respect to hydrogen, $\textrm{Q} / \textrm{H}$ as
\begin{equation}\label{e_q_over_h}
    \textrm{Q} / \textrm{H} \equiv A(\textrm{Q}). 
\end{equation}

Classically, there are a number of reasons solar and stellar spectroscopists scale abundances to hydrogen. In terms of observations, stellar abundances are determined as a ratio between the element line opacity and the continuum opacity determined by H in the stellar photosphere. More broadly, hydrogen is the most abundant element within the universe, making up $\sim$90\% of baryonic material by atom number or $\sim$75\% by mass in the universe. Because the formation of most elements heavier than H and He has increased since the Big Bang (with the exception of Li, Be, and B), those elements compared as a ratio to hydrogen will show the rate of that element's production in time.

With that in mind, it should be noted that geological and/or meteoritic abundances are historically scaled to 10$^6$ Si atoms, since it most abundant element by mole in the Earth and Si is an easy to measure positively charged ion in the Earth's crust \citep{Goldschmidt37}. Fortunately, it's possible to convert between the Si-scaled meteoritic (met) scale and the astronomical H-scaled (astro) scale using:
\begin{equation}\label{e_sol_met}
A(\textrm{Q}) = A(\textrm{Q}_{\textrm{astro}}) = A(\textrm{Q}_{\textrm{met}}) + C\,\,.
\end{equation}
The factor $C$ is determined by taking the average ratio of astronomical abundance (per 10$^{12}$ H atoms) per meteoritic abundance (per 10$^6$ Si atoms) from elements that span a wide range of properties to account for the chemical and physical differences between solar and meteoritic data. We refer the reader to \citet{Lodders19} and \citet[][specifically their \S 3.4.5]{Lodders09}, who provide an excellent explanation of this calculation and walk-through example. \citet{Magg22} also explains an important caveat that converting meteoritic abundances to the H-scale imbues two uncertainty terms: one that is with respect to each abundance measurement (uncorrelated) and another that is associated with the conversion itself (systematic). Overall, this comparison is currently only possible for the Sun and therefore varies depending on the solar abundance measurement. We discuss the most popular solar abundances in more detail in \S \ref{s.solar_abunds}, where we provide the $C$ factor for each dataset, where possible.

\subsection{Scaling to Iron or Other Elements}
For the purposes of comparison, it can be useful to scale elemental abundances with respect to another element, often iron. For example, iron is created from multiple events (such as Type I and Type II supernovae) that occur on different time frames; thus scaling an element to iron enables a more direct understanding of the production site and rate for that element \citep[e.g.][and references therein]{Tinsley80, Timmes95, Kobayashi11, Hinkel14}. We have chosen iron in subsequent equations for clarity and consistency with the literature, but note that in the following subsection all non-hydrogen elements can be substituted for iron in the formalism.

For general element, Q, and iron, Fe, with absolute abundance measurement $\textrm{Q} / \textrm{H}$ $=$ \AQ\, and $\textrm{Fe} / \textrm{H} = A(\textrm{Fe})$ respectively, we define the stellar abundance scaled to iron, $\textrm{Q} / \textrm{Fe}$, as

\begin{equation}\label{e_molar_fraction_to_iron_abundance}
\begin{split}
    \textrm{Q} / \textrm{Fe} &\equiv \textrm{Q} / \textrm{H} - \textrm{Fe} / \textrm{H}  \\
                               &=  A(\textrm{Q}) - A(\textrm{Fe}) \\
                               &= \log_{10}(\textrm{q}) + A(\textrm{H}) - \log_{10}(\textrm{fe}) - A(\textrm{H}) \\
                               &= \log_{10}(\textrm{q}) - \log_{10}(\textrm{fe}) \\
                               &= \log_{10}\big(\frac{\textrm{q}}{\textrm{fe}}\big)\\
                               &= \log_{10}\big(\frac{\eta_{\textrm{q}}}{\eta_{\textrm{fe}}}\big).\\
\end{split}
\end{equation}
Substitutions from Eq. \ref{e_q_over_h} and then Eq. \ref{e_q_abs} show the stellar abundance scaled to iron in terms of the number fractions $\textrm{q}$ \& $\textrm{fe}$. By noting that that total mass of the star is the same for both elements, we also relate to the number densities for Q \& Fe, which are $\eta_q$ \& $n_{\textrm{fe}}$, respectively. As the notation suggests and the calculation confirms, values for $\textrm{Q} / \textrm{Fe}$ are independent of the absolute scale with respect to $\textrm{H}$. 

\subsection{Solar Normalization}\label{s.solarnormmath}
After scaling with respect to hydrogen or any other element, observed stellar abundances, $*$, are often compared with solar abundances, $\odot$. In dex-abundance-notation, a solar normalization is denoted with $[\,\,]$ \citep[which may have been introduced by][]{Helfer59}. We will first consider the case where both solar and stellar abundance values are themselves normalized in reference to hydrogen (i.e. an absolute elemental abundance). We define the solar normalized abundances, $[\textrm{Q}/\textrm{H}]$, as 

\begin{equation}\label{e_solar_abundance_notation}
\begin{split}
    [\textrm{Q}/\textrm{H}] &\equiv \textrm{Q}_{*}/\textrm{H}_{*} - \textrm{Q}_{\odot}/\textrm{H}_{\odot} \\
                    &= A(\textrm{Q})_* - A(\textrm{Q})_{\odot} \\
                    &= \log_{10}(\textrm{q}_{*}) + A(\textrm{H}) - \log_{10}(\textrm{q}_{\odot}) - A(\textrm{H}) \\
                    &= \log_{10}(\textrm{q}_{*}) - \log_{10}(\textrm{q}_{\odot}) \\
                    &= \log_{10}\Big(\frac{\textrm{q}_{*}}{\textrm{q}_{\odot}}\Big). \\
\end{split}
\end{equation}
The relation to number fractions $\textrm{q}_{*}$ \& $\textrm{q}_{\odot}$ is given by a substitution from Eq. \ref{e_q_abs}. As with scaling to hydrogen, solar normalization is also an intuitive subtraction operation in log-space. In this notation, stars that have stellar abundances that are the same as the Sun will have $[\textrm{Q}/\textrm{H}]=0.0$ dex. 

As with absolute abundances, for solar normalized data we may need to examine general element $\textrm{Q}$ with respect to another element. For example with respect to iron, $[\textrm{Q}/\textrm{Fe}]$, it is defined as

\begin{equation}\label{e_solar_iron}
\begin{split}
    [\textrm{Q}/\textrm{Fe}] &\equiv  \textrm{Q}_{*}/\textrm{Fe}_{*} - \textrm{Q}_{\odot}/\textrm{Fe}_{\odot} \\
                               &=  \log_{10}\big(\frac{\textrm{q}_{*}}{\textrm{fe}_{*}}\big) - \log_{10}\big(\frac{\textrm{q}_{\odot}}{\textrm{fe}_{\odot}}\big)\\
                               &=  \log_{10}\Big(\frac{\textrm{q}_{*} \textrm{fe}_{\odot}}{\textrm{fe}_{*} \textrm{q}_{\odot}}\Big). \\
\end{split}
\end{equation}
The substitution from Eq. \ref{e_molar_fraction_to_iron_abundance}, shows how $[\textrm{Q}/\textrm{Fe}]$ relates to the stellar number fractions, $\textrm{q}_{*}$ \& $\textrm{fe}_{*}$, and solar number fractions, $\textrm{q}_{\odot}$ \& $\textrm{fe}_{\odot}$. Here, again, $[\textrm{Q}/\textrm{Fe}]$ = 0.0 dex means that the abundance ratio is the same as that in the Sun. There are many literature sources that have determined solar abundances and may have observed different elements and/or reported varying measured values -- due in large part to differing techniques (see \S \ref{s.solar_abunds}). Calculations of solar abundance, $ \textrm{Q}_{\odot}/\textrm{Fe}_{\odot} = \log_{10}\big(\frac{\textrm{q}_{\odot}}{\textrm{fe}_{\odot}}\big)$, requires that the solar number fractions $\textrm{q}_{\odot}$ \& $\textrm{fe}_{\odot}$ come from the same literature source. It is therefore not recommended that different solar normalizations be mixed together, for example in order to achieve wider element coverage, since it will impart systematic shifts between solar abundance values across measurements.

In addition, it may be useful to alter the solar normalization of an abundance measurement, in order to, for example, have multiple datasets on the same solar baseline for easier comparison \citep{Hinkel14}. In this scenario, we define the original solar normalization as $A(\textrm{Q})_{\odot, 1}$ and the new solar normalization as $A(\textrm{Q})_{\odot, 2}$. The renormalized abundance value, $[\textrm{Q}/\textrm{H}]_{2}$, is defined as
\begin{equation}\label{e_solar_renorm}
\begin{split}
[\textrm{Q}/\textrm{H}]_2 \equiv [\textrm{Q}/\textrm{H}]_1 + A(\textrm{Q})_{\odot, 1} - A(\textrm{Q})_{\odot, 2}
\end{split}
\end{equation}
where $[\textrm{Q}/\textrm{H}]_1$ is the original abundance value normalized to $A(\textrm{Q})_{\odot, 1}$. In other words, the solar renormalization is dependent only on the difference between the two solar normalizations, which is fairly simple when kept in log-space. We go into more depth regarding different solar abundance determinations in \S \ref{s.solar_abunds}.

\subsection{Alternative Notations}
While in this paper we have worked to formalize a consistent mathematical framework that underpins the stellar abundance notation, our choice of notation is not the only one that exists within the literature. Namely, there are a variety of ways to express the same concepts that we have outlined and it is important to understand the alternative notations to promote mathematical comprehension.

For example, while we have defined the absolute abundance of an element in Eqs. \ref{e_q_abs} and \ref{e_q_over_h} as $A(\textrm{Q})$, it may also be equivalently presented in other literature sources as $\log~\epsilon(\textrm{Q})$, $\log(\textrm{Q}$), or $\log(\textrm{Q}_{6158 \textrm{\AA}})$, where the subscript in the latter example specifies the rest-frame, radial-velocity corrected wavelength of the line used to calculate the abundance, or 6158 \AA\, in this case \citep[e.g.][]{Bertran15}. In this notation, [$\textrm{Q}$/H] $\equiv$ $\log~\epsilon(\textrm{Q})$ - $\log_{\odot}~\epsilon(\textrm{Q})$, where the latter term is the absolute value of $\textrm{Q}$ within the Sun \citep[e.g.][]{Shan21}. We also note that, for these alternative notations, all of the $\log$s are implicitly base 10. 

Eq. \ref{e_q_abs} defines the absolute amount of an element $\textrm{Q}$ as the number of $\mathcal{Q}$ atoms per 10$^{12}$ H atoms, which may also be given as log$_{10}$(N$_{\textrm{Q}}$/N$_\textrm{H}$)+12. Similarly, Eq. \ref{e_solar_iron} can then be expressed as [$\textrm{Q}$/Fe] $\equiv$ $\log_{10}$(N$_{\textrm{Q}}$/N$_{\textrm{Fe}})_*$ - $\log_{10}$(N$_{\textrm{Q}}$/N$_{\textrm{Fe}}$)$_{\odot}$ as seen in \citet[e.g.][]{Roederer18} or \citet{Danielski22}.

\section{Mass Fractions}\label{s.massfractions}
For those not steeped in stellar minutia, it is important to point out that the total amount of elements heavier than H and He, or ``metals", varies in stars. For the solar neighborhood ($\sim$500 pc or $\sim$1630 ly from the Sun), most of the stars are between 0.1 -- 3 times the heavy element content of the Sun. There are trends in the ratios of different elements with respect to overall heavy element content that are rooted in how elements are produced by stars over the course of galactic history. For example, the ratio of Mg to Fe tends to go down as the total heavy element content increases. There is also a scatter in addition to those overall trends, such that Fe/Si may vary somewhat at a given heavy element abundance.

Theoretical discussions of stellar evolution and stellar interiors most frequently treat abundances in terms of mass fraction, $X_{\textrm{Q}}$, or nucleon fraction, $Y_{\textrm{Q}}$, which is determined from the mass fraction divided by atomic mass (equivalent to a mole fraction of nucleons). The mass fraction is simply the fraction of the total mass of a given amount of material made up by a given isotope, $X_{\textrm{Q}} = \frac{N_{\textrm{Q}}M_{\textrm{Q}}}{\rho N_A}$, where $N_{\textrm{Q}}$ is the number of atoms of isotope $\textrm{Q}$, $M_{\textrm{Q}}$ is the mass of isotope $\textrm{Q}$, $\rho$ is the mass density of the material, and $N_A$ is Avogadro's number. The difference between mass fraction and nucleon fraction is subtle. If nuclear reactions take place, $\rho$ is not conserved, even if there is no mass flux or change in volume, since some of the rest mass energy of the reactants will be converted into the binding energy of the products (or vice versa in weak interactions and reverse reactions). Nucleon fraction is defined as $X_{\textrm{Q}} = \frac{N_{\textrm{Q}}A_{\textrm{Q}}}{\rho_A N_A}$, where $A_{\textrm{Q}}$ is the atomic mass number. The nucleon density, $\rho_A$, is defined as
\begin{equation}\label{nucleon_dens}
\rho_A \equiv \frac{1}{N_A}  \sum_i A_i N_i \,, 
\end{equation}
where we sum over all elemental (or molecular) species, $i$, in a star. Using the nucleon fraction, as opposed to mass fraction, is more accurate for nuclear network calculations, because it correctly preserves the total mass-energy density and results in a more accurate equation of state. Thus it is typically used for modeling, but in general the effects are small and should never be measurable in the context of observed stellar abundances. For example, the maximum possible difference in mass excess of one atom of $^4 \textrm{He} = 4.86 \times 10^{-28}$g, or 0.7\% of the rest mass of an atom of H. For simulating a particular star, assuming a 1:1 equivalence between mass and nucleon fraction when creating initial model conditions is sufficient. This is why using $X$ interchangeably to denote both mass and nucleon fraction is standard practice. The sum of mass or nucleon fractions $ \sum_i X_i = 1 $. A related quantity is $Y_{\textrm{Q}} = \frac{X_{\textrm{Q}}}{A_{\textrm{Q}}} = \frac{N_{\textrm{Q}}}{\rho N_A }$, the mass fraction divided by atomic weight in AMU, or the ratio of the number of nuclei to the total number of nucleons in the system. $Y_{\textrm{Q}}$ does not sum to one by definition.

A similar and somewhat confusing astronomical notation is $X,Y,Z$, where $X$ is the mass fraction of hydrogen, $Y$ is the mass fraction of He, and $Z$ is the total mass fraction of all other elements, also known as the metallicity. The mass fraction is defined such that $X$+$Y$+$Z$ = 1. This notation is most frequently used in stellar astrophysics theory, because the heavy element content of the star is the second most dominant effect on the stellar evolution (after mass). Historical paucity of multi-element abundance determinations has made the total metallicity, $Z$, the default parameter considered in stellar modeling. Though the evolutionary effects are dependent on the detailed composition, the quoted $Z$ usually implies a specific {\it assumed} set of abundance ratios. The notation is also frequently used in work on stellar populations, where comparisons are made with stellar model grids calculated with fixed $Z$ values and/or observational limitations preclude detailed composition measurements. The metallicity is frequently expressed in two sets of units: one is the direct mass fraction of metals, such that $Z=0.015$ is a mass fraction of heavy elements of 1.5\%; the second gives the metallicity relative to solar, i.e. $Z = 0.1Z_{\odot}$ is one tenth solar metallicity. This distinction is important, as the latter carries the implicit assumption that the proportions of each isotope are the same as they are in the Sun (see \S \ref{s.solar_abunds}, also Table 7 in \citealt{Lodders19} for the $XYZ$ mass fractions from different solar normalizations). In reality, the abundance ratios can vary substantially from solar, and there are infinite combinations that can result in the same $Z$. Different abundance ratios summing to the same $Z$ can have substantial effects on the stellar evolution through modification of the opacity of the stellar material and on the initial composition of a proto-planetary disk. 

It is worth noting at this point that the term ``metallicity" is used observationally in a fashion similar to $Z$. In this case, $[\textrm{M}/\textrm{H}]$ is the dex-abundance-notation equivalent, where $\textrm{M}$ is the sum of all elements heavier than He and follows the same rules as the general case, $[\textrm{Q}/\textrm{H}]$ defined above. It is possible to convert between $[\textrm{M}/\textrm{H}]$ and $Z$ as described below for individual elements with one important caution: since $[\textrm{M}/\textrm{H}]$ is inherently a number fraction and $Z$ a mass fraction, the same $Z$ made up of different proportions of various elements will yields different $[\textrm{M}/\textrm{H}]$. As a simple example, it would take many more O atoms to make up 1\% of a star's mass than it would Fe atoms, meaning O would have a higher number fraction than Fe for the same mass fraction. Therefore converting directly from $[\textrm{M}/\textrm{H}]$ to $Z$ or vice versa requires choosing an assumed set of elemental abundance ratios; converting element-by-element for as many species as are measured gives a more precise result. A final complication is that the overall metallicity and $[\textrm{Fe}/\textrm{H}]$ are often used interchangeably. This is primarily a historical convenience, since up until the last $\sim$20 years, most stars with abundance information only had measurements of $[\textrm{Fe}/\textrm{H}]$. It was therefore necessary to adopt a set of abundance ratios to address any other elements. Most often it was assumed that all elements scaled the same, such that the scaling of Fe would give the scaling for the total metallicity. It has become clear, however, that abundance ratios can vary substantially between stars with the same $[\textrm{Fe}/\textrm{H}]$, so $[\textrm{M}/\textrm{H}]$ must be treated as a distinct quantity. Still, it is common to see references in the literature discuss a star's ``metallicity" when they are really only referring to the star's iron content, [Fe/H].

The simplest conversion from mass fraction to an observationally reported unit of measurement is the absolute abundance (also called atom number) or $A(\textrm{Q})$, as defined in Eq. \ref{e_q_abs_abundance}. However, even this first step is not straightforward. It is best to begin with the \YQ = \AQ\ $- \log_{10}(f_\textrm{H})$, where the normalization to hydrogen is $f_H = \sum_iN_iA_i$, or the sum of atom number times atomic number for all species in a star. But complications enter into the normalization to $A(\textrm{H})$ = 12 (see \S \ref{s.hnorm}). It is tempting to calculate \YQ\, for a star with a given [$\textrm{Q}$/H] as simply \YQ = [$\textrm{Q}$/H] $+$ \AQ\sol - 12. Changing the amount of any heavy element, however, changes the relative number of of hydrogen atoms and therefore the normalization. Ideally $f_\textrm{H}$ would be recalculated for each star by summing all of the absolute measured abundances, but this is rarely practical  given how sporadically elements are measured within stars. Failing this, the common strategy is to assume  some increase in total metallicity and assume a coproduction rate of He with the total metals, or $Y/Z$, to get the new normalization. Estimates of the $Y/Z$ production ratio are scattered and come from a wide variety of sources, including eclipsing binaries, late type field stars, and HII regions \citep{Ribas00, Raul03, Casagrande07, Izotov07}\footnote{Additional references can be found here: \\ \url{www.pas.rochester.edu/~emamajek/memo_dydz.html}.}, with a median value of 2.1 from literature determinations. For a star with a metallicity difference from solar, we see that $\Delta Z = Z-Z$\sol\, and therefore $Y = Y$\sol $+ 2.1\times \Delta Z$. It follows that $X = 1 - Z - Y$, from which the normalization can be determined as above. The best compromise available in most cases is to sum the measured elements, which under ideal circumstances constitute most of Z, and use the empirical $Y/Z$ and an assumed scaling of unmeasured heavy elements with measured species to determine $f_\textrm{H}$. We provide a more tangible example of these calculations for HIP 544 in \S \ref{s.examplecase}.

\section{Stellar Abundances to Molar Ratios}\label{s.abundstomoles}
It is often more useful for fields outside of astronomy to use the stellar elemental abundance information in the form of moles or molar ratios. Therefore, we must step out of log-space and also remove the solar comparison implicit in the dex-abundance-notation. To convert the original measurement of $[\textrm{Q}/\textrm{H}]_*$ to moles Q$_*$, we use:
\begin{equation}\label{e_simple_moles}
\begin{split}
\textrm{Q}_* &= 10^{\textrm{Q}_{*}}  \\
&= 10^{\textrm{Q}_{*}/\textrm{H}_*}, \,\,\,\, \textrm{then from Eq. \ref{e_solar_abundance_notation}}\\ 
&= 10^{([\textrm{Q}/\textrm{H}]_* + \textrm{Q}_{\odot}/\textrm{H}_{\odot})} \\
&= 10^{([\textrm{Q}/\textrm{H}]_* + \textrm{Q}_{\odot})}\, , 
\end{split}
\end{equation} 
where $\textrm{Q}_{\odot}$ is the solar value used to normalize $\textrm{Q}$.

In addition, while comparing $\textrm{Q}$ to H and Fe are common techniques in the stellar abundance sub-field to understand the chemistry of a star in the context of galactic chemical evolution or with respect to the composition of stellar populations, other ratios may also be informative. Therefore, we introduce $\textrm{R}$ as the hydrogen normalized stellar abundance of another element, where $[\textrm{R}/\textrm{H}]_*$ is the solar normalized abundance within the star for the $\textrm{R}$ element. Now we are able to define $\textrm{Q}/\textrm{R}$ as the molar ratio of $\textrm{Q}$ and $\textrm{R}$ elements in the star, such that:
\begin{equation}\label{e_molar_fraction}
\begin{split}
\textrm{Q}_*/\textrm{R}_* &\equiv \frac{10^{\textrm{Q}_*}}{10^{\textrm{R}_*}} \\
                          &= \frac{10^{\textrm{Q}_{*}/\textrm{H}_*}}{10^{\textrm{R}_{*}/\textrm{H}_*}},\,\,\,\, \textrm{then from Eq. \ref{e_solar_abundance_notation}}\\
                            &= \frac{10^{([\textrm{Q}/\textrm{H}]_* + \textrm{Q}_{\odot}/\textrm{H}_{\odot})}}{10^{([\textrm{R}/\textrm{H}]_* + \textrm{R}_{\odot}/\textrm{H}_{\odot})}}\\
                            &= \frac{10^{([\textrm{Q}/\textrm{H}]_* + \textrm{Q}_{\odot})}}{10^{([\textrm{R}/\textrm{H}]_* + \textrm{R}_{\odot})}}\,,
\end{split}
\end{equation}
where $\textrm{R}_{\odot}$ is the solar value for the $\textrm{R}$ element that was specifically used in the original measurement of $[\textrm{R}/\textrm{H}]_*$. We discuss the various solar normalizations standard in the stellar abundance community and how they compare in \S \ref{s.solar_abunds}.

\subsection{Propagating Errors to Moles}
To determine the appropriate error propagation, we start with the uncertainty of any function of one variable $x$, or $f(x) = y$. From \citet[][their Eq. 3.23]{Taylor97}, given that $x$ is measured with uncertainty $\delta x$, then the uncertainty in $y$, called $\delta y$, is given by: 
\begin{equation}\label{e_general_error_func_frac}
\delta y = \Bigl|\frac{dy}{dx}\Bigr| \,\delta x\,\,,
\end{equation}
where $\Bigl|\frac{dy}{dx}\Bigr|$ is the absolute value of the derivative of $f(x)=y$. We note that $\delta x$ is half of the total uncertainty of $x$ or the plus/minus error of $x$, and thus it's a value that is non-zero and positive.

The way that errors are reported for stellar elemental abundances will vary depending on the methodology and/or team who determined the measurements. For example, errors may be reported on a star-by-star basis, i.e. ``the carbon abundance for HIP 100 is [C/H] = 0.11 $\pm$ 0.06 dex" or from Eq. \ref{e_general_error_func_frac}, $\delta$x = 0.06 dex. Alternatively, a general error may be given for an element across all stars, i.e. ``we estimate a total uncertainty in [N/H] of 0.08 dex," or $2 \,\delta$x = 0.08 dex. In the latter case, it's assumed that the error distribution is centered around the measurement such that the total uncertainty is equal to plus/minus half the total uncertainty, or $\pm$ 0.04 in our example. However, propagating errors in the dex-abundance-notation isn't fully applicable because it is inherently in log-space and normalized such that the Sun $\equiv$ 0.0 dex.

For error propagation, we have to work in the linear space, namely Eq. \ref{e_simple_moles}. Since we need to calculate a derivative, we start by redefining Eq. \ref{e_simple_moles} as a more general power law: $y = 10^{(x + c)}$, where $y = \textrm{Q}_*$, $x=[\textrm{Q}/\textrm{H}]_*$, and a constant $ c=\textrm{Q}_{\odot}$. Now we can rewrite $\textrm{Q}_*$ in terms of the natural number $e$, and the natural logarithm $\ln()$:
\begin{equation}\label{e_power_law}
\begin{split}
y &= \textrm{Q}_*([\textrm{Q}/\textrm{H}]_*)=\textrm{Q}_*(x) \\
&= 10^{(x + c)} \\
&=10^{x} \times 10^{c} \\
\Longleftrightarrow \ln(y) &= \ln(10^{x} \times 10^{c}) \\
&= \ln(10^{x}) + \ln(10^{c})\\
&= x\,\ln(10) + c\,\ln(10)\\
\Longleftrightarrow y &= e^{x\,\ln(10) + c\,\ln(10)}\, . \\
\end{split}
\end{equation}

\noindent
From here we can calculate the derivative of the power law as:
\begin{equation}\label{e_der_power_law}
\begin{split}
\frac{dy}{dx} &= \frac{d(e^{x\,\ln(10) + c\,\ln(10)})}{dx} \\
&= \ln(10)\, e^{x\,\ln(10) + c\,\ln(10)} \\
&= \ln(10)\, y
\end{split}
\end{equation}

\noindent
Applying the results of Eqs. \ref{e_power_law} \& \ref{e_der_power_law} to Eq. \ref{e_general_error_func_frac} we are able to express fractional uncertainty  $\frac{\delta y}{|y|}$, i.e. dividing the $\delta$ error by the measurement value, as a function of $\delta x $:
\begin{equation}\label{e_general_error_func}
\begin{split}
\delta y &= \Bigl|\frac{dy}{dx}\Bigr| \,\delta x\\
&= | \ln(10)\, y | \,\, \delta x\\
&= \ln(10)\,\, |y| \,\, \delta x \\
\Longleftrightarrow \frac{\delta y}{|y|} &= \ln(10) \,\, \delta x \\
\frac{\delta \textrm{Q}_*}{|\textrm{Q}_*|} &= \ln(10) \,\, \delta [\textrm{Q} / \textrm{H}]_{*}. \\
\end{split}
\end{equation}

\noindent
Given the definition of $\delta$[Q/H]$_*$, which is half the total uncertainty, Eq. \ref{e_general_error_func} will provide a plus/minus fractional or percentage error (see Step 6 of \S \ref{s.examplecase}). We also see that the errors aren't affected by the choice of solar normalization, since the solar terms are no longer part of the equation. Meaning, if we say that we have [$\textrm{Q}$/H] = 0.0 $\pm$ 0.05 dex, that is the same as saying A($\textrm{Q}$) = 7.46 $\pm$ 0.05. Therefore, the errors would be the same for [$\textrm{Q}$/H] and A($\textrm{Q}$). 

Taking this a step further, we propagate the errors for two elements in dex-abundance-notation, namely Eq. \ref{e_molar_fraction}. The standard method for propagating errors is to convert the total uncertainty to fractional uncertainty for both elements in the numerator (num) and denominator (denom) and then add those fractional uncertainties in quadrature: $\sqrt{(\delta_{\textrm{num}}/\textrm{num})^2 + (\delta_{\textrm{denom}}/\textrm{denom})^2}$, per \citet[][their Eq. 3.47]{Taylor97}. We therefore add the error in quadrature in order to get the fractional error of the molar ratio Q/R:

\begin{equation}\label{e_error_QR}
\begin{split}
\delta \frac{\textrm{Q}_*}{\textrm{R}_*}&= \sqrt{  \Big( \frac{\delta \textrm{Q}_*}{|\textrm{Q}_*|}  \Big)^2 +  \Big( \frac{\delta \textrm{R}_*}{|\textrm{R}_*|} \Big)^2}\\ 
 &= \sqrt{  \Big(  \ln(10)  \delta [\textrm{Q}/\textrm{H}]_* \Big)^2 +  \Big( \ln(10)  \delta [\textrm{R}/\textrm{H}]_* ] \Big)^2}\\
&= \sqrt{  \ln(10)^2 \Big(\delta[\textrm{Q}/\textrm{H}]_*^2 + \delta [\textrm{R}/\textrm{H}]_*^2 \Big) }\\
&= \ln(10)  \sqrt{\delta[\textrm{Q}/\textrm{H}]_*^2 + \delta [\textrm{R}/\textrm{H}]_*^2 }.
\end{split}
\end{equation}
We note that while Eq. \ref{e_error_QR} is nearly identical to Eq. 6 presented in \citet[][since $\ln$(10) $\approx$ 2.303]{Hinkel18}, the results from Eq. 6 were treated as full uncertainties in moles when in reality the equation calculates fractional or percentage errors. As a result, the error values provided in Table 5 of  \citet{Hinkel18} were all overreported. Therefore, it is our intention that the detailed walk-through presented here that resulted in Eq. \ref{e_error_QR} will supersede what was presented in \citet{Hinkel18}.

\section{Example Case}\label{s.examplecase}
To provide an example walk through for many of the equations we've discussed, we will examine HIP 544 (HD 166), a G8V-type star with $T_{eff} = 5400 \pm 100$K \citep{2020ApJ...902....3L,Soubiran16}. We use the abundance values reported in the {\it Hypatia Catalog} for our calculations \citep{Hinkel14}, since the Hypatia Catalog is the largest database of stellar abundances for stars within the solar neighborhood, currently containing 80 elements within $\sim$10,000 stars as compiled from +230 literature sources. Note that these values are used merely for performing an example calculation; no endorsement of the measurement from any particular source is implied. We will convert the stellar abundances of HIP 544 to the mass fraction of heavy elements, or $Z$ notation, in Steps 1 and 2 then examine how changes in the abundance of one element affects $Z$ (Step 2 Caveat). Next we will take the individual stellar abundances and renormalize to a different solar normalization (Step 3), then convert the denominator of the stellar abundances to be  with respect to Fe (Step 4). In Step 5 we convert some of the abundance ratios to molar ratios and then propagate the errors in Step 6. The results of these example calculations, where applicable, are given in Table~\ref{tab.hip544}.

\vspace{3mm}
\noindent
\underline{Step 1:} The most straightforward way to calculate the $Z$ mass fraction is to first convert from $[\textrm{Q}$/H] to A($\textrm{Q})_*$.  Eq.~\ref{e_solar_abundance_notation} provides the conversion from $[\textrm{Q}$/H] to A($\textrm{Q})_*$, {\it assuming a particular solar composition relative to which $[\textrm{Q}/\textrm{H}]$ is measured}. The example abundance data, $[\textrm{Q}$/H], is shown in Column 2 of Table~\ref{tab.hip544}, which uses the \citet{Lodders09} solar composition for normalization -- listed in Column 3, where the conversion to A($\textrm{Q})_*$ is provided in Column 4.

\begin{table*}[t!]
\begin{center}
\begin{tabular}{lccccccccrr}
\hline
$\textrm{Q}$ & [$\textrm{Q}$/H]$_\textrm{L09}$ & $A(\textrm{Q})_{{\odot},\textrm{L09}}$ & $A(\textrm{Q})_{*}$ & $X_{\textrm{Q}}$ & $Y_{\textrm{Q}}$ & $A(\textrm{Q})_{{\odot},\textrm{G07}}$ & [$\textrm{Q}$/H]$_\textrm{G07}$ & [$\textrm{Q}$/Fe]$_\textrm{G07}$ & $\textrm{Q}$/Mg$_\textrm{\,G07}$ & $\textrm{Q}$/Mg$_\textrm{\,L09}$ \\
\footnotesize{(1)} & \footnotesize{(2)} & \footnotesize{(3} & \footnotesize{(4)} & \footnotesize{(5)} & \footnotesize{(6)} &\footnotesize{(7)} & \footnotesize{(8)} & \footnotesize{(9)} & \footnotesize{(10)} & \footnotesize{(11)} \\
\hline
\hline
C & ~0.23 & 8.39 & 8.62 & $3.25\text{e-}3$ &  $2.71\text{e-}4$  & 8.39 & 0.23 & ~0.03 & 9.333 & 9.333 \\\relax 
N & -0.07 & 7.86 & 7.79 & $5.61\text{e-}4$ &  $4.01\text{e-}5$  & 7.78 & 0.01 & -0.19 & 1.380 & 1.380 \\\relax 
O & ~0.09 & 8.73 & 8.82 & $6.87\text{e-}3$ &  $4.30\text{e-}4$   & 8.66 & 0.16 & -0.04 & 14.80 & 14.80 \\\relax 
Ne & ~0.10 & 8.10 & 8.20 & $2.08\text{e-}3$ &  $1.03\text{e-}4$  & 7.84 & 0.36 & ~0.16 & 3.548 & 3.5481 \\\relax 
Na & ~0.08 & 6.29 & 6.37 & $3.50\text{e-}5$ & $1.52\text{e-}6$     & 6.17 & 0.20 & ~0.00 & 0.05248 & 0.05248 \\\relax 
Mg & ~0.11 & 7.54 & 7.65 & $7.06\text{e-}4$ & $2.90\text{e-}5$    & 7.53 & 0.12 & -0.08 & -- & -- \\\relax 
Al & ~0.08 & 6.46 & 6.54 & $6.09\text{e-}5$ & $2.25\text{e-}6$     & 6.37 & 0.17 & -0.03 & 0.07762 & 0.07762 \\\relax 
Si & ~0.17 & 7.53 & 7.70 & $9.15\text{e-}4$ & $3.26\text{e-}5$     & 7.51 & 0.19 & -0.01 & 1.122 & 1.122 \\\relax 
S & ~0.11 & 7.16 & 7.27 & $3.88\text{e-}4$ &  $1.21\text{e-}5$   & 7.14 & 0.13 & -0.07 & 0.4169 & 0.4169 \\\relax 
Ca & ~0.19 & 6.31 & 6.50 & $8.24\text{e-}5$ & $2.06\text{e-}6$    & 6.31 & 0.19 & -0.01 & 0.07079 & 0.07079 \\\relax 
Sc & ~0.17 & 3.07 & 3.24 & $5.08\text{e-}8$ & $1.13\text{e-}9$    & 3.17 & 0.07 & -0.13 & 0.00003890 & 0.00003890 \\\relax 
Ti & ~0.20 & 4.93 & 5.13 & $4.20\text{e-}6$ &  $8.77\text{e-}8$   & 4.90 & 0.23 & ~0.03 & 0.003020 & 0.003020 \\\relax 
V & ~0.10 & 3.99 & 4.09 & $4.07\text{e-}7$ & $8.00\text{e-}7$    & 4.00 & 0.09 & -0.11 & 0.0002754 & 0.0002754 \\\relax 
Cr$^*$ & ~0.00 & 5.65 & 5.65 & $1.69\text{e-}5$ & $3.25\text{e-}7$    & 5.64 & 0.01 & -0.19 & 0.01000 & 0.01000 \\\relax 
Mn & ~0.09 & 5.50 & 5.59 & $1.51\text{e-}5$ & $2.90\text{e-}7$    & 5.39 & 0.20 & ~0.00 & 0.008710 & 0.008710 \\\relax 
Fe & ~0.19 & 7.46 & 7.65 & $1.62\text{e-}3$ & $2.90\text{e-}5$    & 7.45 & 0.20 & -- & 1.000 & 1.000 \\\relax 
Co & ~0.15 & 4.90 & 5.05 & $4.30\text{e-}6$ & $7.29\text{e-}8$    & 4.92 & 0.13 & -0.07 & 0.002512 & 0.002512 \\\relax 
Ni & ~0.14 & 6.22 & 6.36 & $8.74\text{e-}5$ & $1.49\text{e-}6$    & 6.23 & 0.13 & -0.07 & 0.05129 & 0.05129 \\\relax 
Cu & -0.01 & 4.27 & 4.26 & $7.52\text{e-}7$ & $1.18\text{e-}8$    & 4.21 & 0.05 & -0.15 & 0.0004074 & 0.0004074 \\\relax 
Zn & ~0.04 & 4.65 & 4.69 & $2.08\text{e-}6$ & $3.18\text{e-}8$    & 4.60 & 0.09 & -0.11 & 0.001096 & 0.001096 \\ 
\hline
\hline
\end{tabular}
\caption{Example case for HIP 544. All abundance references were provided by \citet{Maldonado15}, with the exception of [N/H] which came from \citet{Brewer16} and [Ne/H] which was assumed to be co-produced with [O/H] and [Mg/H], as discussed in the text. The `L09' subscript indicates values calculated using the \citet{Lodders09} solar normalization while `G07' represents \citet{Grevesse07}. $^*$The average value between [Cr I/H] and [Cr II/H] provided by \citet{Maldonado15} was used for [Cr/H].}\label{tab.hip544}
\end{center}
\end{table*}

\vspace{3mm}
\noindent
\underline{Step 2:} Calculating $Z$ requires summing the absolute atom number of all of the isotopes in order to correctly account for the normalization. To do this, we use Eq. \ref{nucleon_dens} to determine $f_H$, such that $\sum_iN_iA_i = \sum_i(10^{A(\textrm{Q}_i)_{*}})A_i$. 

The majority of elements, both measured in the Sun and on the Periodic Table in general, are not measured in a typical stellar abundance survey. In the best case, the unmeasured elements are low abundance species that will not change $f_H$ significantly. For example, $A(\textrm{Li}) = 3.33$ in the Sun meaning that omitting Li from the calculation of $f_H$ would change its value in the 8th significant digit. In contrast, omitting C where $A(\textrm{C}) = 8.39$ renders $f_H$ incorrect in the third significant figure. Therefore the precision of the normalization of H depends on what fraction (by mass or atom number) of the total heavy elements are measured. Lacking a value for an abundant element like C, N, or O thereby introduces significant uncertainty into the $A(\textrm{Q})$ of other elements. For a precision of 1 in $10^{-4}$, using elements with $A(\textrm{Q}) > 5$ in the Sun would generally be sufficient.

The star HIP 544 provides some examples of the difficulties when determining the accurate metallicity without assuming fixed scaling to solar abundances, due to missing element measurements. Since \citet{Maldonado15} do not report [N/H], the value provided in Table~\ref{tab.hip544} is from a different survey than the rest of the elements (which was normalized to the same solar abundance). Abundance determinations from different surveys frequently display systematic differences larger than the quoted errors, so values are not always interchangeable \citep{Smiljanic14, Hinkel16}. Since N is one of the most abundant elements, a measured value is still to be preferred to an arbitrary value, even if there may be some systematic difference between survey techniques, hence the use of \citet{Brewer16}. In addition, there is no observational determination of [Ne/H] since the only available Ne lines are chromospheric lines in the UV. But neon is too abundant to be neglected. We have the option of simply using the solar value (see different measurements discussed in \S \ref{s.solar_abunds}), looking for a star of similar composition with a [Ne/H] measurement, or assuming it was produced in similar proportions to elements at least partially co-produced in nucleosynthetic contexts. For illustrative purposes, we will use the last option, taking an intermediate value between [O/H] and [Mg/H] of [Ne/H] = 0.10 dex. The abundances derived from neutral and singly ionized Cr lines differed (see \S \ref{s.ions}), both provided by \citet{Maldonado15} and were within errorbars, so we chose to use the median value of the two.

The biggest complication in calculating $f_{\textrm{H}}$ is the change in He abundance. Helium is very rarely measured in stars, so we assume co-production with metals, with $\Delta Y = 2.1\Delta Z$ as discussed in \S \ref{s.massfractions}. Taking $\sum_{i>{\textrm{He}}} (10^{A(\textrm{Q}_{i})_{*}} ) A_i $, the sum increases by 32\% over solar. Using the \citet{Lodders09} value for $Z_{\odot} = 0.0154$, we can estimate that $\Delta Y = 2.1(0.32\times0.0154) = 0.0104$. Ideally, $Y$ would be calculated iteratively with the renormalized values of \YQ\, for the heavier elements, using this value of $Y$ as a starting point for the calculation. (Similarly for the  lacking N value, one could scale N by a factor $f\Delta Z$ as we have done with He, since both are produced primarily in CNO cycle H burning. Analogous strategies can be used for other elements.)

For HIP 544, the resulting $f_{\textrm{H}} \sim 12.187$. For the \citet{Lodders09} solar composition, $f_{\textrm{H}} \sim 12.149$. For each element, \YQ = $A(\textrm{Q}) - \log_{10} f_{\textrm{H}}$, and mass fraction \XQ = \YQ$\times A_{\textrm{Q}}$. Mass fraction \XQ\, is typically quoted in literature instead of \YQ\, as it is more intuitive. Mass fractions \XQ\, are shown in Column 5 of Table~\ref{tab.hip544}. However, \YQ\, is inherently produced by the conversion and is useful for nucleosynthetic calculations, as discussed in \S~\ref{s.massfractions}. The \YQ\, values are shown in Column 6 as calculated from \XQ/A$_{\textrm{Q}}$ using the atomic weights on the standard periodic table. The total metallicity of the star $Z \sim$ 0.0184.

An additional item of note is that the values of $A(\textrm{Q})_{\odot}$ used in this example are the recommended {\it primordial} solar values of \citet{Lodders09}, not the present day photospheric values (\S \ref{s.solar_abunds}). Since HIP 544 is a young star, the measured abundances should reflect the bulk abundances of the star fairly closely. Such is not the case for the average few gigayear old field star. The abundances of He and heavier elements in the photospheres of stars decrease relative to their total stellar abundance over time. Gravitational settling results in heavier elements sinking towards the core of a star. This is partly counteracted by radiative levitation, in which momentum of photons captured by atomic transitions drives atoms outwards. Both processes depend on atomic number Z. The picture is further complicated by advective processes (e.g. convection, rotation, and internal waves) that transport material from one part of the star to another. The settling process takes place on long timescales, so the effect is minimal for massive stars or young low mass stars. In the Sun, the present photospheric $Z$ is $\sim 10$\% less than the bulk solar composition. For the purposes of stellar evolution calculations and inferences about the composition of protoplanetary building blocks, the bulk/primordial composition of the star is the relevant quantity. In order to derive primordial compositions from measurements, stellar models with adjusted initial compositions must be run iteratively until the predicted photospheric abundances match the observed values at the current stellar age.

\vspace{3mm}
\noindent
\underline{Step 2 Caveat:}\, In many cases, $Z$ is assumed to scale with [Fe/H], which is often a necessary assumption because only the [Fe/H] abundance is known. But variations in the [$\textrm{Q}$/Fe], even on the level seen in the solar neighborhood, can have dramatic effects. To illustrate, we will use O as an example element (with a note that the process applies to all species) given the measured quantities [O/H] = 0.09 dex, $A(\textrm{O})_{\odot}$ = 8.73 dex, and $A(\textrm{O})_{*}$ = 8.82 dex. The primary effect on stellar evolution from changing metallicity is due to the change in opacity of stellar material as heavier elements with more electron transitions are added. Since O is the most abundant element after He, changing its abundance can have substantial evolutionary effects. Analysis using individual, self-consistent surveys to remove systematic scatter show that [O/Fe] at a given [Fe/H] can vary by more than a factor of 2 in either direction. To show the effect on $Z$ from this level of variation, we recalculate the $Z$ of HIP 544 with the abundance of O increased by a factor of 2, keeping other species the same and renormalizing to H. In this case, since $\log_{10}(2) \sim$ 0.30, then $A(\textrm{O})_{*}$ = 9.12 dex. The new value gives $f_\textrm{H}= 12.191$, $X_O = 0.0154$, and $Z = 0.0260$. This is a very substantial change in $Z$ with only a tiny change in [Fe/H] (due to the change in the number of H atoms). A star with $Z$ = 0.026 because of an enhancement in O will evolve differently from a star with $Z = 0.026$ because of an enhancement in Fe, since Fe provides more opacity per gram than O. Planet formation would similarly be different for the two cases due to the difference in oxidation state and stoichiometry of mineral building blocks. Therefore using $Z$ (or [Fe/H]) as a fundamental parameter without specifying the detailed composition is theoretically inadequate, even if often necessary. When given the option, it is always preferable to use the specific abundances measured within a star, especially those from a variety of nucleosynethetic origins.

\vspace{3mm}
\noindent
\underline{Step 3:} Instead of considering the heavy elements in bulk, we now walk through our HIP 544 example with respect to the individual elements to convert to a different solar normalization. The abundances provided in the Column  2 of Table \ref{tab.hip544} are normalized to the \citet{Lodders09} solar measurement. Using Eq. \ref{e_solar_renorm}, we can renormalize the measurements with respect to the \citet{Grevesse07} solar measurement, which is provided in Column 7. The renormalized [$\textrm{Q}$/H]$_\textrm{G07}$ abundances are shown in Column 8.

\vspace{3mm}
\noindent
\underline{Step 4:} The second step of Eq. \ref{e_molar_fraction_to_iron_abundance} shows us how to compare $\textrm{Q}$ to an element other than H and change the denominator, for example [$\textrm{Q}$/Fe] = [$\textrm{Q}$/H] - [Fe/H]. We perform this operation on Column 8 of Table \ref{tab.hip544} in order to produce Column 9, [$\textrm{Q}$/Fe]$_\textrm{G07}$. We note that the [Fe/Fe] value was left blank. 

\vspace{3mm}
\noindent
\underline{Step 5:} We switch now from the dex-abundance-notation to calculate molar ratios using Eq. \ref{e_molar_fraction}, where Mg is the comparison element, or $\textrm{R}$ term. Continuing with the data that is normalized to \citet{Grevesse07}, we use Column 8 of Table \ref{tab.hip544} as [$\textrm{Q}$/H]$_*$ and Column 7 for $A(\textrm{Q})_{\odot}$, where [Mg/H] = 0.12 dex is [$\textrm{R}$/H]$_*$ and $A(\textrm{Mg})_{\odot}$ = 7.53 dex is $A(\textrm{R})_{\odot}$. We calculate the molar ratios with respect to Mg in Column 10, $\textrm{Q}$/Mg$_\textrm{\,G07}$, where the Mg/Mg entry was left blank. To show that the molar ratios are irrespective of solar normalization, we perform the same calculation again, but using Columns 2 and 3, respectively, where [$\textrm{R}$/H]$_*$ = [Mg/H] = 0.11 dex and $A(\textrm{R}_{\odot})$ = A(Mg$_{\odot}$) = 7.54 dex. The $\textrm{Q}$/Mg molar ratios calculated from the \citet{Lodders09} solar normalization are shown in Column 11, where we see that they are identical to the calculation from \citet{Grevesse07} in Column 10.

\vspace{3mm}
\noindent
\underline{Step 6:} Propagate molar ratio error. If the typical total error for $\delta$[O/H] = $\pm$0.08 dex and $\delta$[Mg/H] = $\pm$0.06 dex, as determined by \citet{Maldonado15}, then use can use Eq. \ref{e_error_QR} to propagate the error. Namely,  $\delta$O/Mg = ln(10)\,$\sqrt{0.08^2+0.06^2}$ = 0.23 or 23\%, such that the molar ratio O/Mg = 14.80 $\pm$ 3.40 dex.

\vspace{3mm}
\noindent
\underline{Equation Takeaway}: In our example case for HIP  544, we ended up using a variety of the equations presented in this paper. To consolidate and make the calculations more accessible, we provide  Table \ref{tab.eqtakeaway} that shows each step (column 1) with a description of the tasks performed (column 2), the main definitions and equations that were used (column 3), as well a list of alternate notations and additional notes, where applicable (column 4).

\begin{deluxetable*}{p{0.4cm}p{3.0cm}p{7cm}p{4.0cm}}
\caption{Equation takeaway from the \S \ref{s.examplecase} HIP 544 example}\label{tab.eqtakeaway}
\tablehead{ \colhead{Step} & \colhead{Description} & \colhead{Relevant Equations} & \colhead{Alternate Notations \& Notes} 
}
\startdata
1 & Convert from [Q/H] to A(Q$_*$) using the solar composition & A(Q) $\equiv$ absolute stellar abundance to H &  $\log~\epsilon(Q)$, $\log(Q)$, $\log(Q_{\lambda})$, $\log_{10}(N_{Q}/N_H$)+12  \\
 &  & [Q/H]  $\equiv$  Q$_{*}$/H$_{*}$ - Q$_{\odot}$/H$_{\odot}$ =  A(Q)$_*$ - A(Q)$_{\odot}$, solar normalized abundance to H  & [$Q$/H] $\equiv \log~\epsilon(Q) - \log_{\odot}~\epsilon(Q)$  \\
\hline
2 & Calculate the Z mass fraction & $\rho_A \equiv \frac{1}{N_A}  \sum_i A_i N_i\,$, nucleon density for species\, i,  &  \\
    &  & q $\equiv$ number fraction, & \\
      &  & h  $\approx$ 1, \,number fraction for hydrogen & \\
  & & X$_{Q}$ = Y$_{Q}$ $\times$ A$_{Q}$, mass or nucleon fraction & \\
\hline
3 & Renormalize to different solar composition & [Q/H]$_2$ = [Q/H]$_1$ + A(Q)$_{\odot, 1}$ - A(Q)$_{\odot, 2}$ & There are dozens of solar abundance determinations.\\ 
 &  & & \\
\hline
4 & Change abundance ratio denominator  & Q/Fe $\equiv$ A(Q) - Fe/H,\, absolute normalized to Fe &  \\
  &  & [Q/Fe]  $\equiv$ Q$_{*}$/Fe$_{*}$ - Q$_{\odot}$/Fe$_{\odot}$ = [Q/H] - [Fe/H]& [$Q$/Fe] $\equiv$ $\log_{10}$(N$_{Q}$/N$_{Fe})_*$ - $\log_{10}$(N$_{Q}$/N$_{Fe}$)$_{\odot}$ \\
\hline
5 & Calculate molar ratios & Q$_*$/R$_*$ = $\frac{10^{([\textrm{Q}/\textrm{H}]_* + \textrm{Q}_{\odot})}}{10^{([\textrm{R}/\textrm{H}]_* + \textrm{R}_{\odot})}}$ &  \\
\hline
6 & Propagate errors to molar ratios & $\delta  \frac{\textrm{Q}}{ \textrm{R}} = \ln(10) \sqrt{\delta[\textrm{Q}/\textrm{H}]_*^2 + \delta [\textrm{R}/\textrm{H}]_*^2 }$  & Determines fractional (percent) error. Using half of the total $\delta$ errors yields half ($\pm$) the total $\delta$Q/R error.\\
\enddata
\end{deluxetable*}

\section{Varying Solar Abundances}\label{s.solar_abunds}
Because the Sun is the nearest and therefore most studied star, it is a useful first rung in a ladder to understand the elemental abundances of other stars, as shown in \S \ref{s.solarnormmath}. However, the chemical comparisons involve a number of assumptions and caveats that may subtly change the true understanding of a star's composition. While we refer the interested reader to excellent reviews of the Sun's composition by \citet{AllendePrieto20, Lodders19}, we provide a short overview here.

While we can spectroscopically observe the Sun and measure the photospheric elemental abundances, it is possible to verify the abundances of most elements within the Sun by comparing them to the composition of other Solar System objects, such as meteorites, moons, asteroids, et cetera. Meteorites in particular, especially carbonaceous CI-chondrites, are expected to have originated from planetesimals that were not super-heated to melting temperatures and therefore reflect the pristine conditions of the average composition within the Solar System \citep{Lodders03, Lodders09}. The meteoritic abundances, with the exception of those elements that are extremely volatile such as H, He, C, N, O, and the noble gases, are an extremely useful dataset which complement and allow the verification of spectroscopically derived abundances from the Sun's photosphere.

Due in part to the validation of solar abundances to meteoritic composition, which is not possible for any other star, the Sun is used as a comparison against which the elements within other stars in the galaxy are compared. However, this comparison presumes that the Sun is a typical star with a composition that is similar to other G-type stars of similar age within the solar neighborhood. However, this assumption may not be entirely true given studies by \citet[e.g.][]{Melendez09, Ramirez09, Bedell18} who found that the Sun is rich in volatile elements (which have a condensation temperature T$_C <$ 900 K, such as C, N, and O) and deficient in refractory elements (T$_C >$ 900 K, such as Na, P, and Si) -- which may have been segregated into planets.

In addition, it is also expected that because the solar abundances are so often measured that they are also well measured, namely, that different solar abundance studies agree. Currently, the Hypatia Catalog \citep{Hinkel14} features $\sim$75 individual solar abundance determinations -- from dedicated studies of the Sun, to line-by-line differential analyses, to those that determine their solar normalization by using the spectra of asteroids such as Ceres or Vesta. The variation between all of these different solar abundance measurements can be quite large. For example, the range across all Hypatia Catalog solar normalizations for two of the most commonly measured elements, Si and Fe, are $\Delta$0.28 and $\Delta$0.26 dex, respectively, in comparison to typical errors which are $\pm$0.05 and $\pm$0.04 dex, respectively. So, it is important to the overall abundance study which solar normalization is being employed (see \citealt{Hinkel16, Jofre15} for a deeper explanation as to other possible discrepancies).

In comparison, the range for the Si, Fe, and most other elements between datasets that specifically focus on the composition of the Sun is much smaller. This is because solar abundance studies require a huge amount of effort and is a dedicated sub-field within the stellar abundance community. Various teams\footnote{In light of some of the topics/authors discussed here, we would like to acknowledge our support of a harassment free research environment. We strongly believe in, and actively take steps to ensure, an open, inclusive community that allows researchers at any stage of their career and from all backgrounds to be welcome and supported.} often publish solar abundance determinations, only to make improvements on them a few years later. In an effort to inform non-experts on the differences between standard solar normalizations so that they may determine which is best for their research application, we provide a brief overview of the most common solar abundance normalizations used within the literature (in chronological order):\\
\underline{\citet{Anders:1989p3165}}: While not the first solar normalization ever published, this is one the oldest normalizations still in modern use. It's favor may be in part because the solar abundances were compiled from abundances found within CI chondrites, the solar corona, and solar photosphere -- which were predominantly from \citet{Grevesse84a, Grevesse84b}. For Eq. \ref{e_sol_met}, their $C$ = 1.554 $\pm$ 0.020.
\vspace{1mm}

\noindent
\underline{\citet{Grevesse:1998p3102}}: This is an additional compilation of solar abundances, where the photospheric values were derived within \citet{Grevesse96} and have notably larger uncertainties compared to the meteoritic data. While element abundances from the solar photosphere and meteorites all agree to within the (large) error, it was noted that ``the solar photosphere is never \textit{at fault}. Past errors have been shown to be due to errors in atomic or molecular data." With that in mind, the continued use of \citet{Grevesse:1998p3102} may be attributed to more accurate line transition probabilities, which markedly reduced the differences between abundances from the Sun and CI-chondrites.
\vspace{1mm}

\noindent
\underline{\citet{Lodders03}}: One of the most highly cited solar abundances in the last +30 years, this solar abundance compilation not only provides a variety of photospheric abundances, but also compares the abundances from 5 different CI-chondrites and computes their weighted average compositions. The final recommended values for the solar normalization (their Table 1) carefully combines both the photospheric and meteoritic data. In addition, they provide the composition of the proto-Sun and early Solar System, i.e. from 4.55 $\times$ 10$^9$ years (Gya) ago. For Eq. \ref{e_sol_met}, their $C$ = 1.539 $\pm$ 0.046.
\vspace{1mm}

\noindent
\underline{\citet{Asplund:2005p7415}}: While the previous photospheric measurements of the Sun were determined using 1D hydrostatic models, this was the first comprehensive 3D hydrodynamic model of the solar atmosphere. As a result, it substantially changed the C and O photospheric abundances, which were considerably higher than those found in meteorites, and brought them into much better agreement. Unfortunately, the updated model also resulted in some issues with respect to other elements (namely, Na, Al, and Si) and created issues matching to helioseismology data and standard solar models that represent the evolution of the Sun.
\vspace{1mm}

\noindent
\underline{\citet{Grevesse07}}: This solar abundance measurement is an extension of the previous \citet{Grevesse:1998p3102} determination but using the 3D hydrodynamic model referenced in \citet{Asplund:2005p7415}. The authors noted, though, that the updated abundances were predominantly the result of updated atomic data and more a realistic treatment of non-local thermodynamic equilibrium (or NLTE, see \S \ref{s.nlte}), as opposed to the improved 3D model of the solar atmosphere. For example, they were better able to match helioseismology data. Interestingly, despite the fact that this solar abundance measurement is $\sim$15 years old, it has gained popularity as a solar normalization within the literature in the last $\sim$5 years.
\vspace{1mm}

\noindent
\underline{\citet{Asplund:2009p3251}}: The abundances determined for the Sun in \citet{Asplund:2009p3251} are notably the most commonly used values for solar composition, easily making up $\sim$50\% of the solar normalizations for abundance measurements within the Hypatia Catalog. Improvements in the solar model as well as atomic and molecular data resulted in better agreement for the refractory elements between the solar photosphere and the CI-meteorites. Differences with helioseismology data and the standard solar model were also greatly improved.
\vspace{1mm}

\noindent
\underline{\citet{Lodders09}}: These solar abundances are an update from their earlier measurement where new data from meteorites, the solar photosphere, and theoretical models are included. Most notably, the photospheric solar abundances were improved and are distinctly lower compared to \citet{Lodders03}. In addition, developments in instrumentation allowed for new trace elements to be measured within the CI-chondrites. Combined recommended solar abundances for the present-day and at the beginning of the Solar System (4.56 Gya) are provided. For Eq. \ref{e_sol_met}, their $C$ = 1.533 $\pm$ 0.042. 
\vspace{1mm}

\noindent
\underline{\citet{Lodders19}}: Provided here are updated solar abundance values, including from CI-chondrite data and a He estimate determined from helioseismology, as well as an extremely thorough review of solar elemental abundances. In addition, abundances determined from meteoritic compositions, CI-chondrites, the solar photosphere, solar corona/wind, overall Solar System abundances, and mass fractions are compared between all major literature measurements to-date, including references discussed here and more. 
\vspace{1mm}

\noindent
\underline{\citet{Asplund21}}: A continued extension of previous solar abundance determinations from the same first author, there is a useful comparison between their different papers in their Table 1. The biggest change introduced in this paper is the greater accounting for full 3D NLTE dynamics (see \S \ref{s.nlte}) for a wide variety of elements. However, it was noted that the ``modelling problem -- a persistent discrepancy between helioseismology and solar interior models...remains intact with our revised solar abundances." See also \citet{Magg22} for a thorough deduction of NLTE solar abundances (i.e. C, N, O, Mg, Si, Ca, Fe, and Ni) using the most up-to-date atomic/molecular line data, NLTE model atoms, and varying solar model atmospheres.

\section{Discussion \& Additional Considerations}\label{s.discussion}
Relating the composition of the host star to the interior structure and mineralogy of a planet is a complicated task, involving a variety of physical and chemical processes. The endeavor is made more difficult by the need to bridge interdisciplinary fields and the realization that different scientific applications benefit from presenting data in a variety of ways. Stellar spectroscopists normalize data to universal quantities, such as H, while comparing the results to our nearest star. In contrast, geochemists, petrologists, and mineral physicists who study the interior chemistry, mineralogy, and phase equilibria of rocky planets normalize abundances to the most prevalent element in the Earth's crust, i.e. Si, and refer to abundances in terms of molar ratios or weight percent oxides. 

It is equally important to create a common lexicon for interdisciplinary data notation, as well as to make the raw data products accessible from stellar abundance astronomers so that geochemists/geophysicists have the required context to understand complex exoplanet geochemical cycles. An excellent biologically-driven example is phosphorus, which is fundamental to DNA, RNA, and ATP and therefore required to be on the planet's surface in some significant amount, yet it has only been measured in $\sim$260 stars \citep{Hinkel20, Maas22}. Therefore, in order to model the most influential planetary rocks and minerals, stellar spectroscopists should focus their attention on measuring stellar elemental abundances that are necessary for building and modeling planet interiors: ``Major Elements:" O, Mg, Al, Si, K, Ca, and Fe; ``Minor Elements:" C, Na, P, S, and Ni; and key contributors to radioactive heat production (or their daughter products): $^{40}$K, $^{232}$Th, and $^{235, 238}$U \citep{Hinkel18}. 

To aid in further clarification of influential nuances within the stellar abundance field, we thought it necessary to discuss the standard practice of measuring lines from different ionization states (\S \ref{s.ions}) and explain the use of stellar models that depart from LTE (\S \ref{s.nlte}).

\subsection{Ionization States}\label{s.ions}
It is easier to achieve a more robust abundance determination from a star's spectrum when there are numerous elemental (or molecular) lines within the observed wavelength range. A variety of lines means an observer is more likely to identify strong lines that are free from blends, temperature-sensitivity, Zeeman-splitting, and other issues, which will yield consistent line-by-line results. While many lines may not always be available, spectroscopists will often examine neutral as well as ionized lines of a given element. 

An element is ionized when it has either gained an electron and has a negative charge (anion) or lost an electron  and has a positive charge (cation). In chemistry, when an element can form multiple cations, they are indicated by roman numerals, where Cu$^{2+}$ is identified as Cu II. However, in the same way that astronomers have bogarted the definition of ``metals," they, too, have refashioned this notation such that a neutral Ti atom is signified as Ti I, while Ti II ``means the singly ionized atom, i.e., the atom with one electron entirely removed" \citep{Millikan24} or Ti$^+$. Similarly, though not as common, lines may also be doubly (e.g. Ti III) or triply (e.g. Ti IV) ionized. While potential confusion is understandable with this specific notation, especially to those from outside the field of astronomy, it is clear that historical precedent is at fault. 

Regardless of the notation, comparing the results from neutral and ionized lines can be informative. Many stellar abundance techniques vary the stellar parameters (T$_{\textrm{eff}}$, $\log$ g, and [Fe/H]) within their models in order to compare the abundances from Fe I and II lines \citep{Jofre15, Hinkel16, Jofre19}. Iterations of the stellar parameters are continued until an ionization and excitation equilibrium is found within the star, or when the the Fe I and II abundances yield the same result. And because it is expected that the neutral and ionized lines will yield similar abundances, many spectroscopists often use a combination of lines from different ionization states when determining the overall abundance of an element. It is for this reason that the Hypatia Catalog denotes an element as [X II/H] when it is computed only from singly-ionized lines, whereas [X/H] could mean that it comes from neutral lines or a combination of neutral and ionized lines.

\subsection{Non Local Thermodynamic Equilibrium (NLTE)}\label{s.nlte}
Many stellar models often assume local thermodynamic equilibrium (LTE) within the interior of the star, where small pockets are considered to be in a well defined steady-state condition that is thermally isolated. However, these  assumptions are not always applicable to the stellar atmosphere and may result in fluctuations in spectral line strengths and changes in overall line shape \citep{Lodders19}. Unfortunately, NLTE calculations require extensive lab data -- such as NLTE atomic models and line transitions for every line of the element \citep{Grevesse07}, as well as computationally expensive stellar atmosphere models to account for the additional physical processes.

While NLTE calculations may be difficult, they are important for resolving major abundance discrepancies. For example, the differences between abundances from the solar photosphere vs chondritic meteorites discussed in \S \ref{s.solar_abunds} were greatly reduced when accounting for NLTE corrections. Effects from NLTE may also result in an ionization imbalance (per \S \ref{s.ions}), or when calculations using neutral and singly ionized lines result in markedly different elemental abundances that cannot be reconciled by varying stellar parameters \citep[e.g.][and references therein]{AllendePrieto:2004p476}. Not only may departures from LTE impact metal-poor or evolved stars, but there are specific elements, such as Li and K, which require large ($>$ 0.4 dex) NLTE corrections to improve their overall accuracy \citep[e.g.][respectively]{Lind09, Zhao16}. 

Finally, we should note that, unlike measurements from different ionization states (\S \ref{s.ions}), it is not recommended that abundances derived from LTE and NLTE calculations be combined (e.g. averaged) to determine an overall abundance \textit{for a single element}. Namely, the physics underlying NLTE calculations is dissimilar enough to LTE that it makes the results separately comparable, but not equivalent. The best practice, therefore, is to treat LTE and NLTE abundance measurements as independent.

\acknowledgments 
NRH acknowledges NASA support from grant \#20-XRP20\_2-0125. She would also like to thank Zack Maas for his help troubleshooting the equations, Christy Till for her useful feedback, Jake Clark for his input, as well as Tatertot and Lasagna for their support. All of the authors would like to thank Joe Schulze and Wendy Panero for their very thorough and helpful comments on the paper. The research shown here acknowledges use of the Hypatia Catalog Database, an online compilation of stellar abundance data as described in \citet{Hinkel14} that was supported by NASA's Nexus for Exoplanet System Science (NExSS) research coordination network and the Vanderbilt Initiative in Data-Intensive Astrophysics (VIDA). The results reported herein benefited from collaborations and/or information exchange within NASA's Nexus for Exoplanet System Science (NExSS) research coordination network sponsored by NASA's Science Mission Directorate.

\newpage
\bibliographystyle{aasjournal}
\bibliography{papers}

\end{document}